\documentclass[a4, 10 pt, conference]{ieeeconf}  

\IEEEoverridecommandlockouts                              

\usepackage[pdftex]{graphicx}

\title{\LARGE \bf
User-adaptive Tourist Information Dialogue System\\with Yes/No Classifier and Sentiment Estimator*
}

\author{Ryo Yanagimoto, Yunosuke Kubo, Miki Oshio, Mikio Nakano, Kenta Yamamoto, Kazunori Komatani%
\thanks{*This work was partly supported by JSPS KAKENHI Grant Number JP19H05692.}
\thanks{SANKEN, Osaka University, Osaka, Japan}%
}

\begin{document}
\bstctlcite{IEEEexample:BSTcontrol}

\maketitle
\thispagestyle{empty}
\pagestyle{empty}

\begin{abstract}


We introduce our system developed for Dialogue Robot Competition 2023 (DRC2023). 
Our system has two characteristics.
First, rule-based utterance selection and utterance generation using a large language model (LLM) are combined.
We ensure the quality of system utterances while also being able to respond to unexpected user utterances. 
Second, dialogue flow is controlled by considering the results of the BERT-based yes/no classifier and sentiment estimator. 
These allow the system to adapt state transitions and sightseeing plans to the user.

\end{abstract}

\section{INTRODUCTION}



We aimed to develop a tourist information dialogue system for DRC2023 \cite{robotcompe_overview2023}, with the goal of achieving high user satisfaction with dialogue and sightseeing plans. 
We developed our system based on the system of Team OS \cite{kubo2022team} for ialogue Robot Competition 2022 \cite{robotcompe_overview_past}, which relied solely on rules to select system utterances and controlled the dialogue flow using a state transition network.

The drawback of the rule-based utterance selection is its limitation in providing only innocuous responses to unexpected user utterances.
We generated system utterances for unexpected user utterances using LLM, thereby avoiding monotonous dialogues.


Accurately recognizing user intentions is crucial for controlling dialogue.
We develop two machine-learning-based intention recognizers: yes/no classifier and sentiment estimator.
Yes/no classifier based on pattern matching cannot cover all possible patterns in advance.
We thus introduced a BERT-based yes/no classifier to expand the range of situations our system could handle. 
Estimating the sentiment of all users with a single sentiment estimator would decrease accuracy because sentiment expression depends on individual users.
Therefore, we trained multiple models dependent on user groups with similar expressive tendencies.
The performance improvement of the yes/no classifier and sentiment estimator allows the system to flexibly adapt state transitions and sightseeing plans to the user.

\section{Characteristics}\label{Characteristics}

\subsection{Combination of rule-based utterance selection and LLM-based utterance generation}\label{rule_llm}

Rule-based systems can only provide safe and generic responses when users make unexpected utterances, although the quality of system utterances is ensured.
We used LLM-generated system utterances for situations that cannot be handled by such rules.
The combination of rule-based utterance selection and utterance generation using LLM has improved evaluations of "satisfaction with dialogue." 


We used both rules and LLM to respond to unexpected user utterances while maintaining qualities of system utterances.
Our system prepares various candidate user utterances in advance for each system utterance and selects a manually generated system utterance if the user response matches any candidate utterance.
When the user response does not match any candidate, our system generates a response using LLM.
By using LLM-generated system utterances, it is possible to conduct dialogues that are difficult to handle by using only rules.
For example, the user can respond flexibly to questions about the introduced sightseeing spots or the user's impressions.
In such situations, it is difficult to anticipate all the user utterances in advance, so LLM is used.
This allows users to resolve their questions about sightseeing spots and avoid dialogues with only simple questions. 
We used the ChatGPT API 3.5-turbo as the LLM.

\subsection{BERT-based yes/no classifier and sentiment estimator for state transitions and sightseeing plan determination}\label{yes/no_sentiment}
To improve the evaluation of "satisfaction with dialogue" and "favorability of the response,"
our system determines the destination of state transition based on the result of the robust yes/no classifier and sentiment estimator to the user utterances as well as user utterance understanding results.
We used BERT to construct the yes/no classifier because it is difficult to prepare all the patterns for the classification.
We trained two sentiment estimators, one for users under 50 years old and the other for users over 50 years old,
since the sentiment depends on the user's age and using a single model for all users would lead to a decrease in the accuracy of the estimator. 
The appropriate estimator is selected according to the user's age.


The yes/no classifier and the sentiment estimator were also used to create a sightseeing plan.
Our system creates a sightseeing plan more suitable for the user, and improves the evaluation of "satisfaction with dialogue" and "appropriateness of dialogue."
Our system introduces three sightseeing spots and asks the user about their interest in each of them.
Our system determines the sightseeing spots that the user is interested in based on the output of the yes/no classifier and selects those sightseeing spots as candidates for the plan.
If our system cannot decide the two spots by the above method, it chooses the one that the user has a positive sentiment of when asked about their interest.

The yes/no classifier takes a system question and a user response as input and output whether the response is yes, no, or other.
This classifier was constructed by fine-tuning a pretrained model of BERT\footnote{https://huggingface.co/cl-tohoku/bert-base-japanese-v2}.
For the training data, we used Hazumi \cite{komatani_hazumi}, a multimodal dataset of dialogues between a dialog agent and a human, and dialogue data obtained from tests of our system.

The sentiment estimator used a Feedforward Neural Network as the model.
The input is an embedded representation of the speech recognition result obtained using a BERT model\footnote{https://huggingface.co/cl-tohoku/bert-base-japanese-whole-word-masking}, and the output is a number between 0 and 1.
The higher the output value, the more the user enjoys the dialogue.
Hazumi was used to train the sentiment estimator.
The user's age, which is used to switch the sentiment estimator, is obtained from the output of the image recognition program distributed by the competition organizer.

\section{IMPLEMENTATIONS}

\subsection{Dialogue Flow}
\begin{figure}
  \centering
  \includegraphics[width=\linewidth]{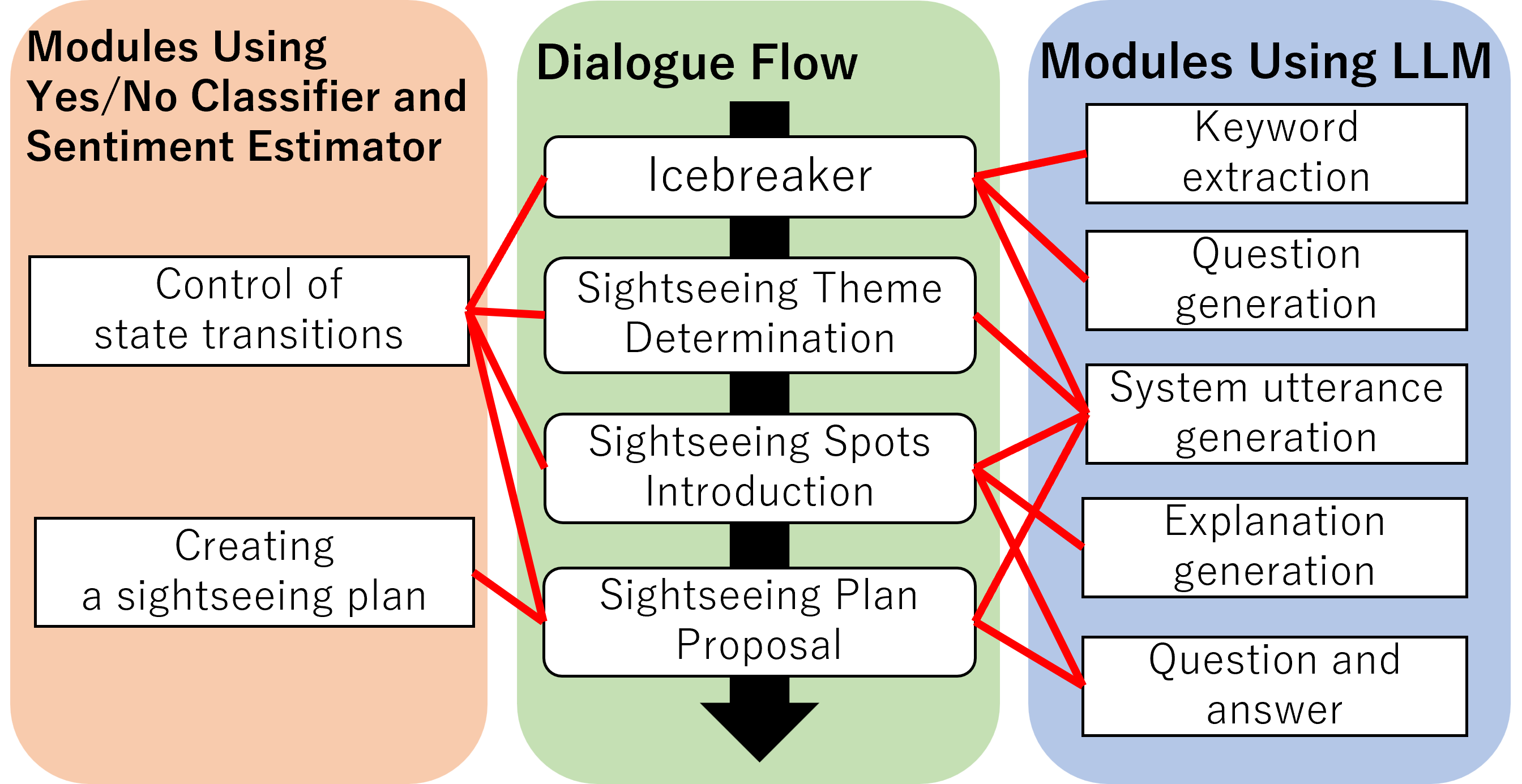}
  \caption{The relationship between dialogue flow and each module}
  \label{dialogue_flow}
\end{figure}

The dialogue flow consists of four phases: "Icebreaker," "Sightseeing Theme Determination," "Sightseeing Spots Introduction," and "Sightseeing Plan Proposal." 
The relationship between the dialogue flow and the modules mentioned in Chapter \ref{Characteristics} is illustrated in Figure \ref{dialogue_flow}.

In the "Icebreaker", casual conversation about sightseeing is done to alleviate the user's tension.
LLM generates questions to dig deeper into the topic if user sentiment is positive. 
Also, we gather useful information to determine the recommended sightseeing spots, such as preferred spots or previously visited spots in Kyoto City.
When users have multiple sightseeing spots they wish to visit, our system asks them where the user most wants to go.
If multiple spots are included in the user response, we extract the spot where the user most wants to go using keyword extraction with LLM.

In the "Sightseeing Theme Determination" phase, we decide on a recommended sightseeing theme for the user. 
We categorized sightseeing spots into three groups: "History," "Nature," and "Others," based on the genres of sightseeing spots available in the database distributed by the competition organizer (hereafter referred to as the "Rurubu DATA").
Through the dialogue, we decide on one theme to recommend to the user.

In the "Sightseeing Spot Introduction" phase, we present three sightseeing spots belonging to the theme determined in the previous phase. 
We avoid introducing sightseeing spots that the user has already visited.
To make a one-day sightseeing plan more feasible, we focus on introducing spots that are geographically close, calculated based on the latitude and longitude obtained from the Rurubu DATA. 
Additionally, we provide explanations for each sightseeing spot generated by LLM, based on the introduction texts retrieved from the Rurubu DATA.
We also accept user questions about each spot and respond using LLM. 
Finally, we inquire about the user's interest in each spot, collecting information to create the sightseeing plan.

In the "Sightseeing Plan Proposal" phase, we propose a sightseeing plan to visit two sightseeing spots.
We create the plan following the methodology outlined in Section \ref{yes/no_sentiment} and explain the reasons for recommending it based on the information obtained during the "Icebreaker."
We include the sightseeing spot that was mentioned as desired during the "Icebreaker" in the proposed plan.
We then engage in a question-and-answer session using LLM regarding the proposed plan and conclude the dialogue.

\subsection{Leveraging DialBB}
We used a state transition network of DialBB\footnote{https://github.com/c4a-ri/dialbb} for dialogue management.
In DialBB, dialogue contents such as states, transition conditions, and rule-based utterances are written and defined on a spreadsheet, so we can separate them from the program.
This allows us to have a clear view of the entire system and to improve and modify easily dialogue contents.

When debugging our system, we used a user simulator with LLM.
By setting up people with different characteristics such as age and gender for the LLM prompt, we were able to collect various dialogue logs and improve the efficiency of development.

\section{CONCLUSIONS}



We combined rule-based utterance selection with utterance generation using LLM to maintain the quality of system utterances and respond to unexpected user utterances. 
We also developed BERT-based yes/no classifier and sentiment estimator, which enabled the system to determine state transitions and sightseeing plans tailored to the user.


\addtolength{\textheight}{-12cm}   








\bibliographystyle{IEEEtran}
\bibliography{IEEEabrv, name}

\begin{thebibliography}{1}
\providecommand{\url}[1]{#1}
\csname url@rmstyle\endcsname
\providecommand{\newblock}{\relax}
\providecommand{\bibinfo}[2]{#2}
\providecommand\BIBentrySTDinterwordspacing{\spaceskip=0pt\relax}
\providecommand\BIBentryALTinterwordstretchfactor{4}
\providecommand\BIBentryALTinterwordspacing{\spaceskip=\fontdimen2\font plus
\BIBentryALTinterwordstretchfactor\fontdimen3\font minus \fontdimen4\font\relax}
\providecommand\BIBforeignlanguage[2]{{%
\expandafter\ifx\csname l@#1\endcsname\relax
\typeout{** WARNING: IEEEtran.bst: No hyphenation pattern has been}%
\typeout{** loaded for the language `#1'. Using the pattern for}%
\typeout{** the default language instead.}%
\else
\language=\csname l@#1\endcsname
\fi
#2}}

\bibitem{robotcompe_overview2023}
T.~Minato, R.~Higashinaka, K.~Sakai, T.~Funayama, H.~Nishizaki, and T.~Nagai, ``{Overview of Dialogue Robot Competition 2023},'' in \emph{Proceedings of the Dialogue Robot Competition}, 2023.

\bibitem{kubo2022team}
Y.~Kubo, R.~Yanagimoto, H.~Futase, M.~Nakano, Z.~Luo, and K.~Komatani, ``{Team OS's System for Dialogue Robot Competition 2022},'' 2022.

\bibitem{robotcompe_overview_past}
T.~Minato, R.~Higashinaka, K.~Sakai, T.~Funayama, H.~Nishizaki, and T.~Nagai, ``Design of a competition specifically for spoken dialogue with a humanoid robot,'' \emph{Advanced Robotics}, vol.~37, no.~21, pp. 1349--1363, 2023.

\bibitem{komatani_hazumi}
K.~Komatani and S.~Okada, ``{Multimodal Human-Agent Dialogue Corpus with Annotations at Utterance and Dialogue Levels},'' in \emph{9th International Conference on Affective Computing and Intelligent Interaction (ACII)}, 2021, pp. 1--8.

\end{thebibliography}

\end{document}